\documentstyle[bbm,amsfonts,epsfig,12pt]{article}

\renewcommand{\a}{\alpha}

\newcommand {\eqref} [1] {(\ref {#1})}
\newcommand {\slsh} [1] {\not{\hbox{\kern-2pt${#1}$}}}

\newcommand {\beq} {\begin{equation}}
\newcommand {\eeq} {\end{equation}}
 \newcommand {\ber}{\begin{eqnarray*}}
 \newcommand {\eer} {\end{eqnarray*}}
\newcommand {\bea}{\begin{eqnarray}}
 \newcommand {\eea} {\end{eqnarray}}
\newcommand{\rmd}{{\rm d}}

\def\Acknowledgements{\bigskip  \bigskip {\begin{center} \begin{normalsize}
             \bf ACKNOWLEDGEMENTS \end{normalsize}\end{center}}}

\begin{document}
\begin{titlepage}
\vskip 1cm
\rightline{CERN-TH/2002-256}
\vskip 3cm
\centerline{{\Large \bf Rotating Strings in Confining AdS/CFT Backgrounds}}
\vskip 1cm
\centerline{{\bf A. Armoni},\,\,{\bf J. L. F. Barb\'on}\footnote{On leave
from Departamento de F\'{\i}sica de Part\'{\i}culas da Universidade de
Santiago de Compostela, Spain.} \,and \,{\bf A. C. Petkou}}
\vskip 0.5cm
\centerline{ \it Theory Division, CERN}
\centerline{\it CH-1211 Geneva 23, Switzerland}
\vskip 0.3cm
\centerline{\it adi.armoni@cern.ch, barbon@cern.ch,  tassos.petkou@cern.ch}
\vskip 1cm

\begin{abstract}

We study semiclassical rotating strings in AdS/CFT backgrounds
that exhibit both confinement and finite-size effects.
The energy versus spin dispersion relation for short strings is the expected
Regge trajectory behaviour, with the same string tension as is
measured by the Wilson loop. Long strings probe the interplay between
confinement and finite-size effects. In particular,
 the dispersion relation for long strings
 shows a characteristic dependence on the string tension and
the finite-size scale.

\end{abstract}
\end{titlepage}

\section{Introduction}

\noindent

The study of  AdS/CFT duality (see \cite{Maldacena:1997re,Gubser:1998bc,Witten:1998qj,Aharony:1999ti})
 beyond the BPS sector has benefited mostly from the semiclassical
approximation.
Any approach of this type is based on two main dynamical
 assumptions: first, we assume
that the AdS-like background that describes the vacuum  can be studied
in the supergravity approximation.
Typically, this involves some  tunable dimensionless number that controls the
 $\alpha'$ corrections in the gravity description. For the standard
example of ${\rm AdS}_5 \times {\bf S}^5$ with radius $R = \sqrt{\alpha'}\,
(4\pi g_s N)^{1/4}$, this quantity is $\alpha' /R^2 = \lambda^{-1/2}$, with
$\lambda = g_{\rm YM}^2 N$ the 't Hooft coupling of the ${\cal N}=4$ super Yang--Mills
 (SYM) theory.
 This type of  condition is common to all AdS/CFT  examples and represents
 the
main obstacle in the application of AdS/CFT ideas to realistic gauge
theories such as QCD.

The second condition of the semiclassical approximation is that the excitation
itself be quasiclassical, in the sense that it can be described as a solitonic
object. This requires some large quantum number that may or may not be of
topological nature \cite{Polyakov:2001af}. Typical examples are the various holographic interpretations
of branes in AdS as instantons, Wilson lines, domain walls, etc.
Another example is the interpretation of the large AdS black hole as a
canonical thermal
ensemble in the CFT. Most of these examples
involve either instantons or static solitonic objects.

 Recently, Gubser, Klebanov and Polyakov (GKP) have studied some instances of {\it stationary}
solitons of periodic type \cite{Gubser:2002tv}.
One particular example where one can go much beyond the semiclassical approximation
is that of highly boosted point-like strings, corresponding to states of
large $R$-charge \cite{Berenstein:2002jq}. Other examples
include truly extended classical string states,
 such as folded strings in the
leading Regge trajectory,
stabilized by a large spin $S$, and/or large  $R$-charge
 quantum number $J$ (see also
\cite{Frolov:2002av}-\cite{Minahan:2002rc} and \cite{deVega:1994yz}
for an earlier work.).

In particular, for very large rotating strings in ${\rm AdS}_5$ in global
coordinates, there is a dispersion relation
\beq\label{drel}
E\,R = S + {\sqrt{\lambda} \over \pi} \;\log \,S\,,
\eeq
valid for $S\gg \sqrt{\lambda} \gg 1$.
 This ``rotor" represents a high-spin state
in the dual CFT defined on a three-sphere  ${\bf S}^3$  of radius $R$. It is
stable in the large-$N$ limit and, by the state/operator map, it corresponds
to a local operator on ${\bf R}^4$. This operator has spin $S$ with respect
to the $SO(4)$ group that leaves the insertion point fixed in ${\bf R}^4$ and
anomalous dimension
\beq\label{anol}
\Delta = E\,R = S+{\sqrt{\lambda} \over \pi} \;\log \,S\,
.\eeq
This is a very interesting prediction that can be compared with similar
results for the anomalous dimensions of twist-two operators for large
spin in weakly-coupled gauge theories such as
realistic QCD \cite{Gross:cs,Floratos:1978ny,Gonzalez-Arroyo:1979he}.

In a previous paper \cite{Armoni:2002xp}, we have shown that the
dispersion relation
(\ref{drel}) applies also in the high-temperature plasma phase,
but only for states of spin $S\gg \sqrt{\lambda} (RT)^4$. This is in agreement
with the expected melting of glueballs when their energy approaches the
average plasma kinetic energy.
In the AdS picture, they correspond to ``planetoids"
\cite{deVega:1996mv,Kar:1997zi} orbiting the large AdS black hole that represents the thermal ensemble.

In this note we study generalizations of the  GKP proposal
\cite{Gubser:2002tv} to backgrounds that describe
confining theories. The appropriate generalization of Eq. (\ref{anol}) is
far from straightforward, because there is no exact one-to-one state/operator
map for non-conformal theories. It was suggested in \cite{Gubser:2002tv}
that non-periodic solitons should be considered in time-dependent backgrounds
(see also \cite{AT}).

 We choose to discuss only
dispersion relations of {\it stationary states} and  postpone the discussion
of what the implications are (if any) for the spectrum of anomalous dimensions of
local operators. Since we are interested in matching the
 dispersion relation (\ref{drel}) at large spin we are led to the consideration
of
AdS models in global coordinates, which represent theories in finite volume.
In section 2 we discuss some general aspects of the interplay between
 confinement and finite-size effects in the context of semiclassical
 AdS/CFT. In particular, in subsection 2.1 we
consider a toy model that illustrates these issues. In section
 3 we study similar features of semiclassical string
 states with large flavour charge in the standard confining model
 of Witten \cite{Witten:1998zw,grossooguri}. Section 4 is devoted to a discussion.

\section{Confinement vs. Finite-Size Effects}

\noindent

In a large-$N$ confining theory in infinite volume where
 glueballs are described by closed-string
states, the obvious candidate for a high-spin soliton is a fast-spinning glueball
in the leading Regge trajectory.  In the large-$N$ limit, it should correspond
to a folded closed spinning configuration of the confining string.
Hence, the classic Regge dispersion relation in flat space,
\beq\label{regge}
E= \sqrt{4\pi\sigma\,S}
\,,\eeq
is a characteristic feature of the glueball spectrum of any
confining theory in the large-$N$ limit. In (\ref{regge}) the constant
$\sigma$ is the string tension, the same as can be measured via the
static potential between heavy quark sources.

It is straightforward to find the required soliton in any AdS/CFT model of
confinement. For a gauge theory in {\it non-compact} ${\bf R}^d$, the
 gravity description  involves, among other things,  a warped
metric
\beq\label{wmet}
ds^2 = f(z) \;ds^2 (\,{\bf R}^d\,) + dz^2 + \dots
\,,\eeq
where $ds^2 (\,{\bf R}^d\,)$ is the standard Minkowski metric on ${\bf R}^d$, the
variable  $z$ represents the holographic coordinate and the dots stand for
internal factors of the full string background. Confinement arises when $f(z)$
has a local minimum  at $z_*$ with $f(z_*) >0$. In many examples,  spacetime
is literally restricted to $z>z_*$. In this situation one has an effective
gravitational potential that stabilizes string configurations at fixed $z=z_*$ and
extended in ${\bf R}^d$. Hence, Wilson lines saturate at $z_*$ with effective
string tension \cite{Maldacena:1998im,Rey:1998ik,Kinar:1998vq}
\beq\label{sten}
\sigma = {f(z_*) \over 2\pi\alpha'}
\,.\eeq
By the same token, folded closed strings that rotate in ${\bf R}^d$ will be
stabilized at $z=z_*$ and satisfy (\ref{regge}) with string tension
(\ref{sten}).

An intriguing result of \cite{Gubser:2002tv} is that semiclassical
Regge trajectories of type (\ref{regge}) can be found in the spectrum of an
exact CFT.
For spins in the range $1\ll S \ll \sqrt{\lambda}$ the folded strings rotate
deep inside AdS and do not feel the curvature. Then one finds the
law (\ref{regge}) with string tension
\beq\label{adst}
2\pi\sigma_{\rm CFT} = {1\over \alpha'} =   {\sqrt{\lambda} \over R^2}
\;.
\eeq
Of course, this is not surprising in itself, given that the model does
contain strings. However, in this case the ``string tension" (\ref{adst}) is tied to  the
scale of  finite-size effects (the size $R$ of the ${\bf S}^3$), and it
cannot be probed by static quark sources because the corresponding quark-antiquark
distance
cannot be larger than   ${\cal O}(R)$.

Hence, we see that standard Regge trajectories can be the landmark of true confining
phenomena, but they can also   appear as artefacts of  finite-size effects at strong
coupling.
In fact, in the GKP description of the ${\cal N}=4$ SYM model, both the
Regge trajectories and the logarithmic correction in (\ref{drel})
arise as peculiar finite-size effects.

It would be interesting to see if true confining Regge trajectories can coexist
with a logarithmic high-spin asymptotics as in  Eq.  (\ref{drel}).
For this purpose, one should consider a confining theory defined on a
spatial 3-sphere of finite radius  $R$,
 with an intrinsic
glueball mass gap $\Lambda_{\rm QCD} \gg 1/R$, so that confining Regge trajectories
can be identified independently of the finite-size effects. If the high-energy
asymptotics, $E\gg \Lambda_{\rm QCD} \gg 1/R$ is well approximated by a
   conformal theory, then one should also find (\ref{drel}) for
very high spin.

\begin{figure}
  \begin{center}
\mbox{\kern-0.5cm
\epsfig{file=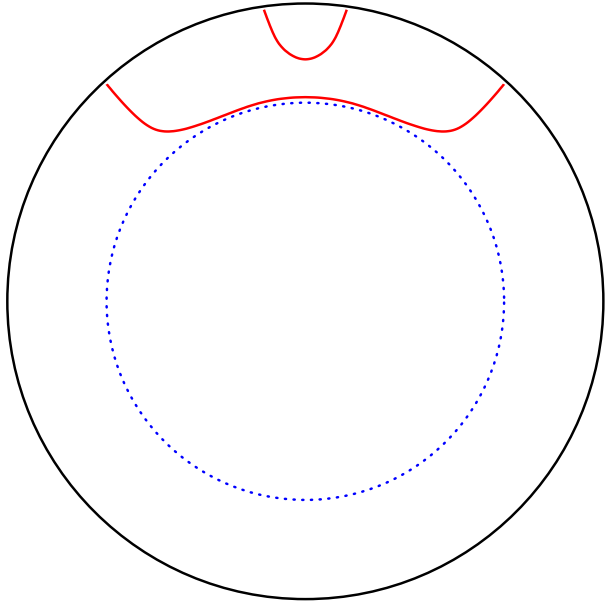,width=8.0true cm,angle=0}}
\label{wilsonfig}
  \end{center}
\caption{{\small Behaviour of Wilson lines in a model of confinement in finite volume.
  The dotted line denotes the ``confining sphere" at  $r=r_*$.  Strings that
correspond to quark-antiquark pairs that are sufficiently separated in the
3-sphere will saturate there.}
  }
\end{figure}

Unfortunately, all existing AdS/CFT models of confinement have the form (\ref{wmet})
and hence describe gauge theories in infinite volume. In order to exhibit a family
of stationary periodic strings  interpolating between confining
 Regge trajectories and
(\ref{drel}), one must construct a background that is asymptotic to ${\rm AdS}_5$ space
in global coordinates,
\beq\label{ass}
ds^2 \approx -(1+r^2 /R^2) \,dt^2 + {dr^2 \over 1+ r^2 /R^2} + r^2 \,d\Omega_3^2 + \dots\,,
\eeq
as $r\rightarrow \infty$, but shows a ``confinement sphere" at some intermediate
radius $r=r_* \gg R$.
 By a confinement sphere we mean a region with a repulsive gravitational
potential for test masses. Under these conditions, static strings that
represents interacting quark-antiquark pairs would behave as in fig. 1. A static string
subtending
 an angle $\theta \ll \pi$ in the angular ${\bf S}^3$ drops to the interior
of AdS down to a  radius of order
$$
r_{\rm min} \sim {R \over \theta}
\,.$$
However, when $r_{\rm min} \sim r_*$, the repulsive gravitational potential stabilizes
the string as in fig. 1. After the standard subtraction of the infinite
``bare quark mass", one is left with a contribution to the interaction energy
that grows linearly with the proper length on the
 spatial ${\bf S}^3$, i.e. a confining static potential.

It is clear from the picture that short folded strings, rotating in a plane
tangent to the 3-sphere at $r=r_*$, will be stabilized from falling
 to smaller $r$,
just like the static strings are. If the length of the rotor is small
with respect to
the radius of the confining sphere, the dispersion relation will be approximately
Regge-like, $E \sim \sqrt{S}$, with some effective string tension. On the other hand,
for very large spin, the rotor is much larger than the confining sphere, and its
properties are well approximated by the rotor in vacuum AdS space (\ref{ass}). In
this way we describe a family of periodic solitons that interpolate between confining
Regge trajectories and the logarithmic behaviour (\ref{drel}).

\subsection{A Toy Model of Confinement in Finite Volume}

\noindent

It would be very interesting to exhibit a consistent AdS/CFT model with the
qualitative features shown in fig. 1. Lacking such a model, we
introduce in this section a formal toy model with the required properties.
It can be constructed as an unphysical
 limit of
an R-charge condensate of ${\cal N}=4$ SYM on ${\bf S}^3$. The AdS/CFT
representation of a thermal ensemble with such non-vanishing charge $Q$
 is in terms of an AdS charged black hole with metric
\cite{Chamblin:1999tk}
\beq
ds^2= -f(r)\, dt^2 + f^{-1}(r)\, dr^2 + r^2 \,d\Omega ^2
\label{global}
\,,\eeq
with
\beq
f(r) = 1+r^2 -{M\over r^2} + {Q^2 \over r^4}.
\label{QM}
\eeq
Note that the metric is asymptotic  to AdS in global coordinates, with units
chosen so that
   $R=R_{\rm AdS}=1$.
For large values of $M/Q$ the equation $f(r)=0$ has two positive roots,
the largest one being the horizon of the black hole. In the extremal limit
the two roots coalesce into a single one. At this point the black hole
temperature vanishes and the solution represents a zero-temperature
charge condensate. The associated extremal mass $M$ gives the energy
cost of such a condensate. If we now ``overcharge" the black hole beyond
 the extremal limit, the function $f(r)$
 becomes positive and develops a minimum
at some positive radius $r_*$.

Since $f(r)$ is a gravitational potential for static masses at fixed
$r$, the 3-sphere at $r=r_*$
 satisfies the conditions for a ``confinement wall", in the sense of
the previous section.
 On the other hand, the singularity at $r=0$ becomes naked
and presumably unphysical (i.e. not resolved by stringy effects). In order
to simplify the formulas, we can take the extreme situation with $M=0$.
 Then $r_* \sim Q^{1/3}$ and
$r_* \gg R_{\rm AdS}$ for  $Q\gg 1$. In this regime,
 the region of the geometry that is responsible for
the  confinement  properties is well separated from the naked
singularity.

Hence, we have a geometry that  {\it simulates} a confining
behaviour  in finite volume, in such a way that the
confining length scale is well contained inside the box.
 We use this setup simply as a formal
model, since the original condensate in the ${\cal N}=4$ SYM
theory is rather unphysical for $M/Q \rightarrow 0$. We think,
nevertheless, that any consistent gravity description of a similar confining theory
in finite volume should have the generic properties exhibited by this
model.

Semiclassical rotors
 of proper length much smaller than $r_*$ stabilize
 in a plane tangent to the ``confining sphere" $r=r_*$, as in
fig. 2. Let us
pick a north pole for this sphere and single out a two-dimensional
subspace of the tangent space at this north pole,  with polar coordinates
$(\rho, \phi)$. Then, in the vicinity of this point, the
relevant (2+1)-dimensional metric  is approximately flat:
\beq\label{apl}
ds^2 \approx f(r_*)\left (\,-dt^2 + d\rho^2 + \rho^2 \,d\phi^2 \right )
\,.\eeq
Hence, there is
an approximate Regge trajectory for small strings stabilized around
the north pole
with the dispersion relation
\beq\label{effs}
E\approx \sqrt{4\pi\,\sigma\,S}\,,\qquad \sigma ={f(r_*) \over 2\pi\alpha'}
\sim {Q^{2/3} \over \alpha'}\;.
\eeq
As the spin grows, these rotors start feeling the curved background
geometry; they will follow the dashed line in fig. 2 and eventually
 their proper length becomes much larger than $r_*$. At this
point we can approximate the dispersion relation of such ``eccentric" rotors
by that of the central rotors that pass through the singularity.

A folded string rotating on
an equator of ${\bf S}^3$ with angular velocity $\omega$  has action
\beq\label{ngac}
I_{\rm NG} = -{2\over \pi\alpha'} \int dt \,dr\,\sqrt{1-{\omega^2\, r^2
\over f(r)}}\,.
\eeq
Therefore, it will be subluminal if
$$
\omega^2 \, r^2 \leq f(r)
\,,
$$
with the equality determining the folding points of the string.
We see that, in the limit that $M\rightarrow 0$, this equation
has solutions of finite energy only for $\omega >1$, and they
correspond to rotors that extend from the naked singularity
at $r=0$ up to the turning point $r_{\rm max}$.

For $Q\gg1$, there are three regimes of interest. First, we have
short strings with $r_{\rm max} \ll Q^{1/3}$. These are presumably irrelevant
to our purposes, since their properties are characteristic of the
region dominated by the singularity. It is interesting, however, to
notice that both the energy and the spin of such strings are finite. A
simple scaling argument yields the law (see Appendix A for an explicit calculation)
$
E^4 \sim Q\,S/ {\a '^3}
$.

Conversely, for very long rotors we expect the dispersion relation
to approach that of standard AdS space, since most of the string
is rotating far from the ``confining" region. So, we obtain
$$
E \sim S + {\sqrt{\lambda} \over \pi} \;\log\,S
$$
asymptotically for very large spin.

The interesting feature of this model is that in addition to the two
above asymptotic regimes there exists an intermediate one where
the rotor endpoints are in the range:
\beq\label{intrange}
Q^{1/3} \ll r_{\rm max} \ll Q^{1/2}\,.
\eeq
In Appendix A we prove that in this regime the rotors obey the
following law
\beq \label{thelaw}
E\,R-S \approx A\, \left ( 1- c\, (A/S)^{ 1/2} \right )
\;,
\eeq
where we have restored the radius of the sphere by dimensional analysis. The
constant $c$ is a positive numerical coefficient of ${\cal O} (1)$ and
\beq\label{defa}
A \propto {Q^{1/3} \,R^2 \over \alpha'} \sim \lambda^{1/4} \,\sqrt{\sigma \,R^2}\;.
\eeq

\begin{figure}
  \begin{center}
\mbox{\kern-0.5cm
\epsfig{file=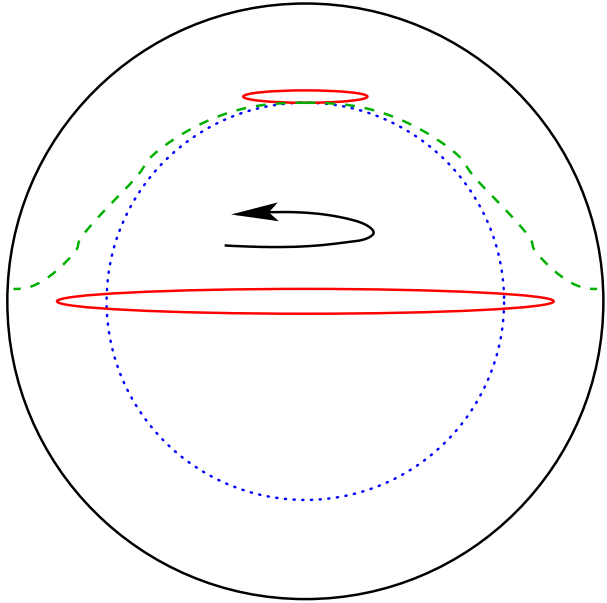,width=8.0true cm,angle=0}}
\label{rotorsfig}
  \end{center}
\caption{{\small Rotors in the charged black hole model. The dotted sphere
  denotes the radius $r=r_* \sim Q^{1\over 3}$. The ``eccentric''
  rotors sit at $r=r_*$, whereas the ``central'' rotors pass through $r=0$.}}
\end{figure}

Therefore, we arrive at the following picture. The flat-space Regge
behaviour characteristic of confinement extends for spins in the
range $1\ll S\ll \sqrt{\lambda}\,Q^{1\over 3}$.
At approximately the upper limit there
 is a crossover
to the particle-like relativistic behaviour
 (\ref{thelaw}). This intermediate regime extends in the range of
spins
\beq\label{interm}
\sqrt{\lambda} \,Q^{1\over 3} \ll S \ll \sqrt{\lambda}\,Q \;,
\eeq
where the upper limit signals the final crossover to the logarithmically
corrected relativistic behaviour that is characteristic of the conformal
theory on ${\bf S}^3$.

The rotors can be considered as probes of the peculiar interplay
between confinement and finite-size effects in these models. One way
of expressing this interplay in graphical terms is to define a notion
of effective length of the string in terms of gauge-theory quantities.
One possible definition would be in terms of the effective world-sheet
area spanned in one period of motion, equal to $2\pi/\omega$.
That is, we write
the Nambu--Goto action as
\beq\label{eflen}
2\pi \alpha' \; I_{\rm NG} = - {2\pi \over \omega} \cdot L_{\rm eff}\;.
\eeq
 With this definition, $L_{\rm eff}$ measures the
correction to the ultrarelativistic dispersion relation $ER=S$, i.e. we have
\beq\label{elav}
L_{\rm eff} =  2\pi \alpha' \,(E-\omega\,S ) \,.
\eeq
A confining Regge trajectory
is characterized by strings that grow linearly with energy
\beq\label{lin}
L_{\rm eff} (E) \propto E \;.
\eeq
On the other hand, the subleading
logarithm of conformal theories on ${\bf S}^3$
gives an anomalous logarithmic growth of the string $L_{\rm eff} \propto
\log (E)$ in this regime. The intermediate regime (\ref{interm}) is
associated to a saturation effect on the length of the string, which stays
``locked" on the size of the box
\beq\label{int}
L_{\rm eff} \approx R\,, \qquad {\rm for} \;\;\;\sqrt{\lambda}\, Q^{1\over 3} \ll S
\ll \sqrt{\lambda} \, Q \;.
\eeq
We illustrate this behaviour in fig. 3.

Hence, we see that the spectroscopy of high-spin glueballs in these
models reveals the finite-size effects in the high energy regime, in
yet another interesting manifestation of UV/IR effects for extended
objects.

\begin{figure}
  \begin{center}
\mbox{\kern-0.5cm
\epsfig{file=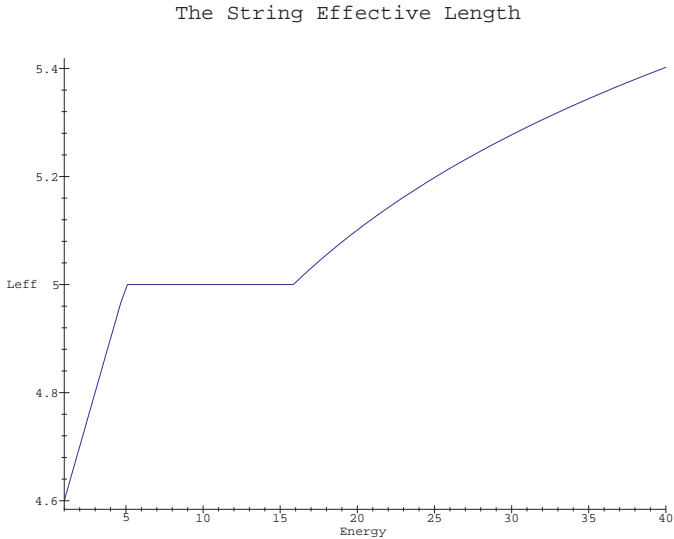,width=8.5true cm,angle=270}}
\label{lengthfig}
  \end{center}
\caption{ \small The effective length of the string as a function of energy.
  For energies lower than $Q^{1\over 3}/ R $ (short strings) the length
  grows linearly. Finite-size effects induce a transient plateau  where the
string length is locked to the size of the box.
  For larger energies this plateau tilts with  the  logarithmic growth characteristic of
the conformal fixed point.}
\end{figure}

\section{Rotating Strings in Witten's Model of Confinement}

\noindent

An entirely different aspect of the interplay between confinement and finite-size
effects is provided by the models in refs.  \cite{Witten:1998zw, grossooguri}.
 In this case, one starts from a non-confining
SYM theory on ${\bf S}^1 \times {\bf R}^p$, with supersymmetry-breaking boundary
conditions on the circle of radius $R$. In the limit of dimensional
reduction, one should recover ordinary non-supersymmetric Yang--Mills
theory on ${\bf R}^p$. Thus, supersymmetry-breaking by finite-size effects in a higher
dimensional theory induces standard confinement (in infinite volume) after dimensional
reduction.  The purpose of this section is to use the semiclassical rotating strings
as probes of this mechanism.

   A large-$N$ gravity description of these models
is obtained by
looking at the metric of near-extremal D$p$ branes \cite{Itzhaki}
\beq
ds^2 = r^{7-p \over 2} \left ( R^2 \,h(r) \,
     d\phi ^2 + d{\vec y}^{\;2} -dt^2 \right ) + r^{p-7 \over 2}
 {dr^2 \over h(r)} + {r^2 \over r^{7-p \over 2}}
\;  d\Omega ^2 _{8-p}
\,,\eeq
where the function $h(r)$ is given by
$$
h(r) = 1-\left({r_0 \over r}\right)^{7-p},
$$
and we use units such that the charge radius of the D$p$-brane is set
to unity: $1=R_q \sim \sqrt{\alpha'} (g_s N)^{1 \over 7-p}$.
The supergravity approximation to this background is good when the
't Hooft coupling of the $(p+1)$-dimensional SYM theory is large at
the scale of the compact circle, i.e. when
$$
g^2_{p+1} \,N \,R^{\,p-3} \gg 1\;.
$$
The metric is smooth at $r=r_0$ if $r_0$ is fixed as a function of
$R$ via
\beq
\label{Rr0}
R={2\over 7-p} \,r_0^{\;{p-5 \over 2}}\;.
\eeq
Hence, space-time terminates at $r=r_0$, which behaves as a ``confining
wall" in this model. The associated string tension, as measured by
static quark sources separated in ${\bf R}^p$ is
\beq\label{sr}
\sigma = {r_0^{\;{7-p \over 2}} \over 2\pi\alpha'}\,.
\eeq
Hence, folded closed strings sitting at $r=r_0$ and rotating in a plane
contained in ${\bf R}^p$ will follow a Regge dispersion
 relation $E= \sqrt{4\pi\sigma\,S}$. These
rotors are unaffected by finite-size effects at high energy because
${\bf R}^p$ is non-compact.

The interesting feature of this model is that stable rotors  also occur
along the ``cigar" factor of the metric, namely the $(r,\phi)$ space
that fills the ${\bf S}^1$ circle in the bulk supergravity background.
These rotors have constant angular velocity in the ${\bf S}^1$ circle
of radius $R$, i.e. $\phi = \omega\,t$. The associated angular
momentum  is nothing but the linear momentum along the compact circle. We
shall denote its quantum by $n$.

The action of these solitons is given by
\beq\label{acw}
I_{\rm NG} = -{2\over \pi\alpha'} \int dt\,dr\sqrt{{1\over h(r)} -\omega^2 R^2}
\;,
\eeq
where the range of the radial variable is $r_0 \leq r \leq r_{\rm max}$. The
turning point is given by the solution of
$$
\omega^2 R^2 = {1\over h(r_{\rm max})},
$$
so that $\omega R$ approaches unity from above when $r_{\rm max} \rightarrow \infty$.
The expressions for the energy and quantized compact momentum of these
solitons are
\bea
& & E =  {2\over \pi\alpha'}  \int _{r_0} ^{r_{\rm max}} {dr \over h(r)} {1\over
 \sqrt { h(r)^{-1} - \omega ^2 R ^2\;,
     }} \\
& & n =  {2\over \pi\alpha'} \int _{r_0} ^{r_{\rm max}} dr {\omega\, R ^2  \over
 \sqrt { h(r)^{-1} - \omega ^2 R ^2 }}
\;.\eea
These integrals are reducible to standard hypergeometric
functions, which we present in Appendix B.

There are two interesting regimes. For
$r_{\rm max} \ll R$ we have short rotors in locally flat space near the tip of
the cigar at $r\approx r_0$. These rotors follow linear Regge trajectories
of the form
\beq\label{linnn}
E=\sqrt{4\pi\sigma\,n}
\,.
\eeq
The second regime corresponds to very long rotors $r_{\rm max} \gg R$ that
``climb up" the cigar towards the boundary. The corresponding dispersion
relation is
\beq\label{findis}
E\,R = n + b\,\sigma R^2 \left[ 1- c\,\left({\sigma R^2 \over n}\right)^{\nu}\; \right]\;,
\eeq
with $b, c$ positive numerical constants and $\nu = (5-p)/(9-p)$. Hence,
we find the same saturation behaviour as in the previous model. The
effective string length $L_{\rm eff}$ grows linearly with the energy
up to the crossover momenta of order $n\sim \sigma R^2$. At these values
of the momentum it levels off  to a plateau that continues for arbitrarily high
energies; the ``finite-size
locking".

The physics of the long ``cigar rotors" in this model is rather
interesting when compared with the supersymmetric counterpart corresponding
to the same ${\bf S}^1 \times {\bf R}^p$ space-time with supersymmetric
boundary conditions on the circle. In that case the circle remains
non-contractible in the bulk, since $r_0 =0$.  The cigar becomes a cylinder
and  string states  of momentum $n$ are point-like in the classical
approximation. The corresponding dispersion relation is $E=n$, up to
corrections of order 1 in string units.

The nonsupersymmetric model behaves very differently. By looking at
the $n\rightarrow \infty$ asymptotics, we would expect to probe just
the short-distance behaviour of
the  supersymmetric $(p+1)$-dimensional theory.
However, we see that solitonic objects remain at any arbitrarily
large momentum in  such a way that their wave function
 has a non-trivial overlap
with all scales down to the confining scale $1/R$.  The leading
deviation from the supersymmetric dispersion relation is semiclassical
and depends on the string tension of the low-energy confining theory.

\section{Discussion}

\noindent

One of the most striking aspects of AdS/CFT models in the strong coupling limit
 is the  emergence of higher-dimensional bulk physics, codified in the Hilbert
space of the four-dimensional CFT. The prime example of this phenomenon is
the density of states of ${\cal N} =4$ SYM on ${\bf S}^3$ (see
 \cite{Aharony:1999ti} for a summary).
For $N\gg \lambda \gg 1$ a large hierarchy of regimes opens up between the naive
mass gap $E\sim 1/R$ and the large-$N$ phase transition threshold $E\sim N^2/R$.  For
example, in the range $1/R \ll E \ll \lambda^{1/4} /R$ the density of states
is well approximated by a massless ten-dimensional gas. At the upper limit the
density of states turns into a Hagedorn spectrum  of ten-dimensional
strings with tension
of order $1/\alpha' \sim \sqrt{\lambda}/R^2$.

One can interpret the GKP results in a similar vein. In the Hamiltonian interpretation
we are looking at the spectrum of very particular states on ${\bf S}^3$, namely
single-trace (glueball) states with maximal spin for a given energy. In the gravity
dual they correspond to classical rotating folded strings.
In view of the previous considerations regarding the density of states, it is
natural that we find a ten-dimensional Regge trajectory with ``string tension"
$\sigma_{\rm CFT} \sim \sqrt{\lambda} /R^2$ in the range of energies $\lambda^{1/4} /R
\ll E \ll \lambda^{1/2} /R $.
However, it is also clear that in this model the string tension is an artefact
of the finite-size effects. At very high energies the dispersion relation reproduces
an anomalous logarithmic growth of the spinning ``glueball".

In this paper we have studied GKP-type periodic solitons in models where true confinement
phenomena coexist with finite-volume constraints. We identify semiclassical rotors
representing glueball states on Regge trajectories $E=\sqrt{4\pi \sigma S}$, where
$\sigma$ is a true string tension that would survive in the infinite-volume limit.
Our main result is the identification of a ``saturation" effect caused by the
finite volume. Namely, the linear  growth of the effective string size along the
Regge trajectory reaches a plateau when the glueball hits the  walls of the box.

Our interpretation of the  dispersion relations in various examples
 suggests that this saturation is a  generic
effect induced by the finite size of either the physical space or the
 internal space associated with a conserved charge ($R$-charge or compact momentum).

In models whose extreme high-energy behaviour is well approximated by a CFT on
${\bf S}^3$, we find that  the effective size of the glueball ends up
showing the logarithmic growth of GKP. Thus, the existence of
the transient plateau in these models  is really a consequence of
 the hierarchy $\sigma \gg \sigma_{\rm CFT}$, between
the true confining string tension and the artificial string tension of an
exact CFT at strong coupling.

Coming from the high-energy side, we could say that the flattening of
 the CFT logarithm is the signal of confinement at low energy.
 We regard this as an interesting
speculation, because the logarithm itself is visible in ordinary perturbation theory
in realistic QCD. At weak coupling, it arises from gluon-exchange effects  in the
computation of anomalous dimensions. Therefore, it is tempting to regard the
plateau as the onset of confinement effects in the sense that ``gluon effects" become
better described as  ``glueball effects".

The main obstacle in the development of these ideas is the lack of an exact correspondence
between energies on ${\bf S}^3$ and conformal dimensions on ${\bf R}^4$, once we are
away from the conformal fixed point. It would be very interesting to find the appropriate
generalization of this fact  that is useful  to our discussion of  confining models.

\Acknowledgements

A.A. would like to thank A. Font, J. Sonnenschein and S. Theisen
 for useful discussions. A.C.P. wishes to thank E. Floratos for
interesting remarks.

\begin{appendix}

\vspace{0.8cm}

{\Large \bf Appendices}

\section{Dispersion relations for rotating strings in the
charged AdS black hole}

We study rotating strings in the background
\eqref{global},\eqref{QM} (with $M=0$, $Q^2 \gg 1$). We use the
prescription of \cite{Gubser:2002tv}. The string
 configuration extends in the radial $r$ direction of space-time and
 rotates along the $\phi$ angle with
 constant angular velocity $\dot{\phi}= \omega$. The Nambu--Goto action
\beq
\label{NG}
I_{\rm NG} = - {1 \over 2\pi \a '} \int \rmd\tau\rmd\sigma
\sqrt{-\det G_{\mu \nu} \partial _\alpha X^\mu \partial _\beta X^\nu}
\eeq
takes the following form
\beq
\label{action}
I_{\rm NG} = - {1 \over 2\pi \a '} \int \rmd \tau \int \rmd \sigma
\sqrt{-G_{rr} \left ({dr\over d\sigma} \right )^2 (G_{tt} +
 G_{\phi \phi} \dot \phi ^2)  }\,.
\eeq
For a metric of the form \eqref{global}, with \eqref{QM} it yields
\beq
\label{NG1}
I_{\rm NG} = -  {4 \over 2\pi \a '} \int \rmd \tau \int
 _0^{r_{\rm max}}\rmd r
 \sqrt{1 - \dot{\phi}^2\frac{r^2}{f(r)}} = -  {4 \tau \over 2\pi \a '}\int
 _0^{r_{\rm max}}\rmd r
 \sqrt{1 - \dot{\phi}^2\frac{r^2}{1+r^2 + Q^2/r^4}} \,.
\eeq
The integration range $0\leq r\leq r_{\rm max}$ is determined by the
 condition that the square root in \eqref{NG1} be real, namely by the
 requirement
\beq
1+r^2 +{Q^2\over r^4} \ge \omega ^2 r^2.
\label{roots}
\eeq

The energy and angular momentum of the string are then given by
\bea
\label{E}
\hspace{-1.4cm} &&E = {2\over \pi \a'} \int _{0}^{r_{\rm max}}\rmd r \frac{1}{\sqrt{1 -
  \omega^2\frac{r^2}{f(r)}}}=
{2\over \pi \a'} \int _{0}^{r_{\rm max}}\rmd r
  \sqrt{\frac{r^4+r^6+Q^2}{r^4+(1-\omega^2)r^6 +Q^2}}\,,\\
\label{Sp}
\hspace{-1.4cm} &&S = {2\over \pi \a'}  \int _{0}^{r_{\rm max}}\rmd r
  \frac{\omega\frac{r^2}{f(r)}}{\sqrt{1 -
  \omega^2\frac{r^2}{f(r)}}}=
 {2\over \pi \a'}  \int _{0}^{r_{\rm max}}\rmd r
  \frac{\omega r^6}{\sqrt{(r^4+r^6+Q^2)(r^4+(1-\omega ^2)r^6 +Q^2)}}\,.
\eea

The condition for real roots, eq.\eqref{roots} leads to three
interesting regimes
\bea
 \mbox {(i).}  &  r_{\rm max} \ll Q^{1\over 3} & \mbox {Short strings.}
 \label{region1} \\
 \mbox {(ii).} &  Q^{1\over 3} \ll r_{\rm max} \ll Q^{1\over 2} & \mbox
 {Intermediate regime.} \label{region2} \\
 \mbox {(iii).} & Q^{1\over 2} \ll r_{\rm max}  & \mbox {Long strings.}
  \label{region3}  
\eea

\subsection{Short strings}

Let us consider first the region \eqref{region1}. When the
strings are very short, one can choose a very fast rotation, such
that $\omega ^2 r_{\rm max} ^6 = Q^2$, for a given fixed $Q$. For this case
\bea
& & E \sim {1\over \a '} \left ({Q\over \omega} \right ) ^{1\over 3}\,
,
\\
& & S \sim {1\over \a'} \left ({Q\over \omega} \right ) ^{1\over
  3}{1\over \omega}\, .
\eea
Note that although we have a singularity at the origin, both the energy
and the spin do not diverge. For the short string we obtain the
relation
\beq
E^4 \sim {Q \over \a ^{'3}} S\,.
\eeq

\subsection{Intermediate regime}

 Next, consider strings in the regime
\beq
r_{\rm max}\ll Q^{\frac{1}{2}}\, .
\label{cond1}
\eeq
In this regime we can approximate the energy (\ref{E}) and angular
momentum (\ref{Sp}) by 
 \bea
 E&=& \frac{2}{\pi\a'}\int_{0}^{r_{\rm max}}\rmd
 r\left(1+\frac{1}{Q^2}r^6\right)^{\frac{1}{2}}\left(1-\frac{\eta}{Q^2}r^6\right)^{-\frac{1}{2}} , \label{Eapprx}
 \\
S&=& \frac{\omega}{Q^2}\frac{2}{\pi\a'}\int_{0}^{r_{\rm max}}\rmd
 r r^6\left(1+\frac{1}{Q^2}r^6\right)^{-\frac{1}{2}}\left(1-\frac{\eta}{Q^2}r^6\right)^{-\frac{1}{2}}.\label{Spapprx}
 \eea
Now, the parameter $\eta=\omega^2-1$ determines the ``length'' of the
strings since 
\beq
r_{\rm max}\approx \frac{Q^{\frac{1}{3}}}{\eta^{\frac{1}{6}}}\, .
\label{rmax2}
\eeq
In particular, as $\eta\rightarrow 0$ we have $r_{\rm max}\gg
Q^{1/3}$. Therefore, under the assumption that $Q$ is very large, we
can consider ``long strings'' in the regime
\beq
\label{cond2}
Q^{\frac{1}{3}}\ll r_{\rm max}\ll Q^{\frac{1}{2}}\, .
\eeq

The integrals (\ref{Eapprx}) and (\ref{Spapprx}) can be explicitly
evaluated with the results:
\bea
\label{E2}
E &=&
\frac{Q^{\frac{1}{3}}}{6}\frac{2}{\pi\a'}\frac{1}{(1+\eta)^{\frac{1}{6}}} 
B\left(\frac{1}{2},\frac{1}{6}\right)
{}_2F_1\left(\frac{1}{6},\frac{7}{6};\frac{2}{3};
  \frac{1}{1+\eta}\right), \\
\label{Sp2}
S &=&
\frac{Q^{\frac{1}{3}}}{6}\frac{2\omega}{\pi\a'}
\frac{1}{(1+\eta)^{\frac{7}{6}}}  
B\left(\frac{1}{2},\frac{7}{6}\right)
{}_2F_1\left(\frac{7}{6},\frac{7}{6};\frac{5}{3};
  \frac{1}{1+\eta}\right).
\eea
It is interesting to note that despite the fact that both expressions
above diverge as $\eta\rightarrow 0$, they never exhibit logarithmic
singularities. Furthermore, the divergences cancel exactly in the
difference $E-S$ and the result is
\bea
 E-S &=&
 \frac{Q^{\frac{1}{3}}}{6}\frac{2}{\pi\a'}
 \left[\frac{\Gamma(\frac{1}{6})
     \Gamma(\frac{1}{3})}{\Gamma(\frac{1}{2})}
   -9\frac{\Gamma(\frac{1}{2})
     \Gamma(\frac{2}{3})}{\Gamma(\frac{1}{6})}\left(\frac{\eta}{1+\eta}\right)^{\frac{1}{3}}+...\right] \nonumber\\
 &\approx& \frac{Q^{\frac{1}{3}}}{6}\frac{2}{\pi\a'}
 \frac{\Gamma(\frac{1}{6})
     \Gamma(\frac{1}{3})}{\Gamma(\frac{1}{2})}
   -\frac{3Q^{\frac{1}{2}}}{2}\left(\frac{2}{\pi\a'}\right)^{\frac{3}{2}} \left( \frac{\Gamma(\frac{1}{2})
     \Gamma(\frac{2}{3})}{\Gamma(\frac{1}{6})}\right)^{\frac{3}{2}}
 \,\frac{1}{S^{\frac{1}{2}}}+...\,\, .
\eea
The dots denote terms that are less important for $S\gg Q^{1/3}/\a'$.

Finally, for long strings in the regime (\ref{region3}), the space is
asymptotically 
AdS and we recover the GKP relation (\ref{drel}).

\section{Results for the Witten model}

The energy and the flavour charge of the Witten model can also be 
explicitly calculated as
\bea
\label{EWitt}
E &=& \frac{4r_0}{2\pi\a'}\frac{[h(r_{\rm max})]^{\frac{1}{2}}}{7-p}
B\left(\frac{1}{2},\frac{1}{2}\right)
{}_2F_1\left(\frac{8-p}{7-p},\frac{1}{2}; 1;h(r_{\rm max})\right) \\
\label{nWitt}
n &=& \frac{4\omega R^2r_0}{2\pi\a'}\frac{[h(r_{\rm max})]^{\frac{3}{2}}}{7-p}
B\left(\frac{3}{2},\frac{1}{2}\right)
{}_2F_1\left(\frac{8-p}{7-p},\frac{3}{2}; 2;h(r_{\rm max})\right)
\eea
For long strings, $\omega R, \,h(r_{\rm max})\rightarrow 1$ and we obtain
\beq
\label{EnWitt}
ER - n =
\frac{Rr_0}{\a'}\frac{2\Gamma\left(\frac{5-p}{7-p}\right)}{
  (7-p)\sqrt{\pi}\Gamma\left(\frac{6-p}{7-p}\right)}
-\frac{Rr_0}{\a'}\frac{(7-p)
  \Gamma\left(\frac{9-p}{7-p}\right)}{(5-p)\sqrt{\pi}
  \Gamma\left(\frac{1}{7-p}\right)}
\,[1-h(r_{\rm max})]^{\frac{5-p}{7-p}}+...\, ,
\eeq
which leads to (\ref{findis}) by virtue of (\ref{sr}) and (\ref{Rr0}). 

\end{appendix}

\end{document}